\documentclass[a4paper]{ESASPCS13Style}
\usepackage{epsfig}

\newcommand{\COBOLD}{{\sc CO$^5$BOLD}}
\newcommand{\pun}[1]{\mbox{\rm\,#1}}

\newcommand{\Teff}{\ensuremath{T_{\mathrm{eff}}}}
\newcommand{\logg}{\ensuremath{\log g}}

\newcommand{\beq}{\begin{equation}}
\newcommand{\eeq}{\end{equation}}

\begin{document}

\title{Status and future of hydrodynamical model atmospheres}

\author{Hans-G{\"u}nter~Ludwig\inst{1} \and\
  Ar\={u}nas~Ku\v{c}inskas\inst{2}}
\institute{Lund Observatory, Box~43, 22100~Lund, Sweden \and
Institute of Theoretical Physics and Astronomy, Go\v{s}tauto 12,
  Vilnius~2600, Lithuania}

\maketitle

\begin{abstract}
Since about 25 years ago work has been dedicated to the
development of hydrodynamical model atmospheres for cool stars (of
A to T spectral type). Despite their obviously sounder physical
foundation in comparison with standard hydrostatic models, their
general application has been rather limited. In order to
understand why this is, and how to progress, we review the present
status of hydrodynamical modelling of cool star atmospheres. The
development efforts were and are motivated by the theoretical
interest of understanding the dynamical processes operating in
stellar atmospheres. To show the observational impact, we discuss
examples in the fields of spectroscopy and stellar structure where
hydrodynamical modelling provided results on a level qualitatively
beyond standard models. We stress present modelling challenges,
and highlight presently possible and future observations that
would be particularly valuable in the interplay between model
validation and interpretation of observables, to eventually widen
the usage of hydrodynamical model atmospheres within the
astronomical community. 
\end{abstract}

\section{Introduction}

Model atmospheres are our most important theoretical tool for the
interpretation of stellar spectra. They are constructed as
time-independent, one-dimensional, plane-parallel or spherically
symmetric configuration in hydrostatic and radiative-convective
equilibrium. In todays standard models for late-type atmospheres the
radiative energy transport is modelled in a highly realistic manner,
while the convective energy transport is usually treated in a rather
simplistic way, assuming mixing-length theory or related
concepts. Since the assumptions behind the construction of such
``classical'' model atmospheres are obviously not fulfilled, for now
about 25 years efforts have been invested to overcome the limitations
of the classical models: the result are ``hydrodynamical'' model
atmospheres which from first principles account for the
time-dependence, and three-dimensional character of the gas flows
primarily related to convection in the surface layers of late-type
stars. They allow for deviations --- not necessarily small ones ---
from the equilibrium conditions assumed in standard models. The
convective energy transport is described by the very nature of
hydrodynamical models quite realistically, the radiative energy
transport with reasonable accuracy. We say ``reasonable'' since the
convective flows exhibit no symmetries and their time-dependence
renders the radiative transfer problem computationally much more
demanding than in standard models.  Some trade-offs in the
descriptions have to be made which limit the level of the achieved
precision. The higher degree of physical realism of hydrodynamical
models leads to a rich set of atmospheric processes which are not
present in standard models: hydrodynamical models harbour waves and
shocks, and can describe the macroscopic transport and mixing of
stellar matter.

Despite the obviously sounder physical foundation of
hydrodynamical model atmospheres their general application has
been rather limited. In order to understand why this is, and how
to progress, we want to review the present status of
hydrodynamical modelling. The paper reflects the outcome of a
discussion between the authors within an ongoing project on the
photometric properties of late-type giant star atmospheres. One
might perhaps say that the authors represent the astronomical
community in micro-format: the first author works mostly
theoretically and has been actively involved in the development of
hydrodynamical model atmospheres, while the second author is
mostly working observationally being a typical ``end-user'' of
model atmospheres. In our discussion questions concerning the
status and future of hydrodynamical model atmospheres were put
bluntly: Generally, has the development of hydrodynamical model
atmospheres been worth the effort? And, concerning their presently
limited general application and usage: Which spectral type can be
modelled? Why are there no model grids available?  Why are there no
ready-to-use libraries of synthetic spectra and colours available?
Can one expect this to happen anytime soon?  What is needed to
drive further developments? In the following we shall try to give
answers or at least make reasonable projections concerning the
questions posed above.  We want to convince the reader that
hydrodynamical model atmospheres are indeed a useful tool for
interpreting and understanding stellar spectra, and that they will
undoubtedly play an increasingly important role in the future.

\section{Probing atmospheres across the HRD}

\begin{figure}[ht ]
\begin{center}
\includegraphics[width=\hsize]{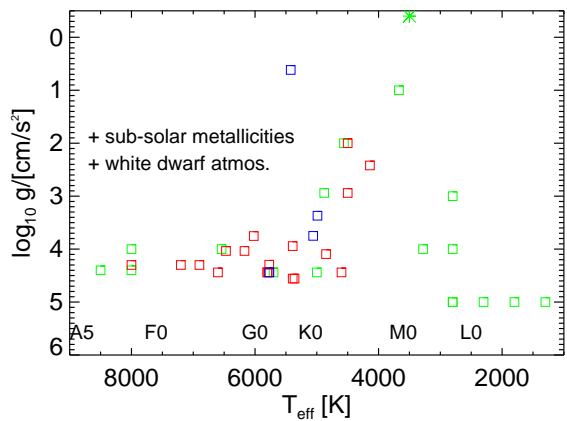}
\caption{Atmospheres probed with 3D hydrodynamical models in the
HRD --- here in effective temperature-gravity plane. Shown are
models of three different research groups (squares; the star denotes a
3D hydrodynamical model of Betelgeuse by Freytag et al. 2002). Not
all models constructed in the groups are depicted, but the
coverage of the parameter space is representative. Note, that also
models of sub-solar metallicity as well as models for white dwarf
atmospheres have been calculated. \label{fig1}}
\end{center}
\end{figure}

We conducted a small survey among three research
groups\footnote{Chan, Robinson, Kim, et al.; Nordlund, Stein,
Asplund, Trampedach, et al.; \COBOLD-group: Steffen, Freytag,
Ludwig, Wedemeyer-B\"ohm, et al.} to obtain an overview which
atmospheric parameters have been probed by hydrodynamical
modelling across the Her\-tzsprung-Russell diagram (HRD). While
not complete we believe that Fig.~\ref{fig1} gives a
representative view of the state of affairs. A sizeable fraction
of the HRD has been covered by 3D hydrodynamical atmospheres, on
the main-sequence stretching from A- to L-type spectra. Naturally,
the Sun and its vicinity has been intensively studied. The bottom
of the red giant branch is probed by several models too, albeit
very sparsely for more evolved objects. Besides the models
depicted in Fig.~\ref{fig1} also a number of hydrodynamical white
dwarf (DA type) model atmospheres has been constructed, and models
of sub-solar metallicity. Some ``uncharted territory'' remains:
e.g. atmospheres of horizontal branch stars, of T-type (or cooler)
brown dwarfs and planets, or of primordial stars.

Many of the models depicted in Fig.~\ref{fig1} are of experimental
nature, and have not necessarily been published yet. Hitherto, the
exploration of the HRD has not been performed in a systematic
manner, but has rather been driven by interests of the various
researchers in particular problems. We nevertheless conclude that
at present most of the atmospheres in the HRD are --- at least in
principle --- accessible to 3D hydrodynamical modelling.

\section{Impact of hydrodynamical model atmospheres}

What kind of astrophysical impact did hydrodynamical model atmospheres
have so far? Perhaps most importantly they led to qualitative and
quantitative progress in our \textit{theoretical understanding\/} of
the atmospheric dynamics and role of inhomogeneities. Hydrodynamical
model at\-mo\-spheres provide information about velocity fields and
thermal inhomogeneities in the layers where the formation of spectral
lines takes place.  Knowledge of this information allows to abandon
the classical free parameters of micro- and macro-turbulence which are
used to describe the line broadening of macroscopic velocity fields in
standard line formation calculations. In this respect hydrodynamical
model atmospheres have reached a level of realism --- in particular in
the case of the Sun --- which allows to reproduce observed spectral
line profiles virtually perfectly without introducing any free
parameters (see e.g. \cite{ALNS00}). As an immediate consequence
abundance analyses are put on a firmer footing, providing
high-fidelity abundances from hydrodynamical modelling.

The realistic description of the convective energy transport in
hydrodynamical model atmospheres allows to abandon another free
parameter entering the calculation of standard stellar model
atmospheres --- the mixing-length parameter. It is introduced
within the framework of mixing-length theory to parameterise the
efficiency of the convective energy transport, and in this way
controls to some extent the temperature gradient in convectively
unstable layers. The temperature gradient influences predicted
spectral properties (general shape of the spectral energy
distribution, spectral line profiles, photometric colours, etc.).

In recent years full-fledged model atmospheres were beginning to
replace more idealised outer boundary condition employed in stellar
structure models (see e.g. \cite{BCAH98}). In particularly this sort
of application demands for a reliable description of the temperature
profile between optically thick and thin layers. Since hydrodynamical
model atmospheres can provide the temperature profile independent of
mixing-length theory a major factor limiting the predictive power of
stellar structure models is removed.

Last but not least, hydrodynamical models can make predictions
about atmospheric phenomena which are connected to time-dependent
processes. E.g. the structure of stellar chromospheres, coronae,
and dust-driven winds is related to wave processes which can only
be described using dynamical models.

In the following sections we shall give a number of examples intended
to illustrate the points above. The results would not have been
possible to obtain with standard model atmospheres. Moreover, they
were selected to show that hydrodynamical models did not only advance
basic theoretical understanding but also had impact on the
\textit{interpretation of observations\/}.

\subsection{High-fidelity abundances}

In a recent paper \cite*{AGNAK04} reported a new determination of
the solar O abundance based on 3D hydrodynamical model
atmospheres together with improved atomic data to obtain
a consistent value of the O abundance from all available
spectroscopic abundance indicators for the first time. The
analysis suggests a significant downward revision of the solar O
abundance, and together with related changes of the abundances of
C, N, and Ne suggests an overall decrease of the solar metallicity
by 35\,\% with respect to the often quoted abundances of
\cite*{AG89}. While the new O abundance is in accordance with the
O abundance of the local interstellar medium, it is in conflict
with results from helioseismology. At the moment it is unclear
where the resolution of this conflict may be found. The conflict
has nevertheless spawned discussions and renewed efforts in the
modelling of the solar structure, nicely exemplifying how progress
in our understanding is most often driven by discrepancies rather
than consistencies. The discrepancy in this particular case became
apparent by the higher precision of the abundance determination
that the application of hydrodynamical model atmospheres made
possible.

A similar example is the work by \cite*{BBA03} where the authors
apply hydrodynamical model atmospheres combined with non-LTE
calculations in the determination of the Li abundance in
metal-poor halo stars. As a result, already existing discrepancies
are aggravated between the Li abundance in old halo stars and the
primordial Li abundance as inferred by the {\sc Wmap} analysis of
the cosmic microwave background.  Here the application of
hydrodynamical models gave confidence that the discrepancy is not
just an artefact due to leaving out granulation induced systematic
effects in the spectral line formation calculations.

This issue has been studied by \cite*{SH02} for the case of the Sun,
but with a wider scope and considering different chemical species \&
spectral lines with different formation properties. The authors set
out to provide upper bounds for the possible errors of spectroscopic
abundance analyses associated with photospheric temperature
inhomogeneities due to granulation. They find in general a
strengthening of lines if temperature inhomogeneities are present. In
this example hydrodynamical model atmospheres have made it possible to
study this problem with a higher degree of realism in the first place.


\subsection{Stellar structure independent of MLT}

Mixing-length theory (hereafter MLT) was introduced into the
theory of stellar structure in the early 1950s (\cite{V53}). It
had the big advantage of providing a simple recipe for calculating
the convective energy flux needed in the model construction.
Simultaneously it introduced (at least) one free parameter --- the
mixing-length parameter --- which had plagued the predictive power
of stellar structure models ever since. The mixing-length
parameter is usually empirically calibrated by comparing the
stellar models with the Sun. The question is, however, whether the
scaling across the HRD provided by the MLT is indeed correct, in
other words, whether the mixing-length parameter is actually
constant as assumed when calculating the evolution of a star in
the HRD. Many investigations have been undertaken to obtain
empirical estimates for the mixing-length parameter over a wider
range of stellar parameters. No clear picture emerged so far. On
the one hand side, many physical effects can produce effects
similar to variations of the mixing-length parameter. On the other
hand, short-comings of the model can also mask or introduce
changes of the mixing-length parameter.

From the viewpoint of stellar structure MLT is mostly important in a
thin boundary layer close to the stellar surface, in this sense
becomes a stellar atmosphere problem. Classical stellar
at\-mo\-spheres themselves rely on MLT, hence can\-not provide new
information. Hydrodynamical model atmospheres, however, can provide an
estimate of the efficiency of the convective energy transport from
first physical principles which can be translated into an equivalent
mixing-length parameter. \cite*{LFS99} used hydrodynamical models to
provide theoretical estimates of the mixing-length parameter for
late-type stars located in the HRD in vicinity of the Sun.

Proper treatment of convection is particularly important in case of
late-type giants (i.e., stars on the red giant and asymptotic giant
branches, RGB \& AGB). Being intrinsically bright they can be
effectively used for tracing distant stellar populations (or those
heavily obscured by interstellar dust), in order to derive properties
of their chemical evolution, star formation histories, etc.\,. A
realistic representation of their interiors and atmospheres by
theoretical models would be indeed crucial in this context, in order
to provide a reliable background for a correct interpretation of the
observables. However, late-type giants, partly because of their very
extended nature, are very complex and correct representation of
dynamical phenomena in their interiors (and immediate vicinity too ---
see Sect.~3.3) relies heavily on the proper treatment of convection, as
even a minor change in the mixing-length parameter may notably shift
the position of a star in the HRD. In their pioneering attempt
to probe the interiors of late-type giants with 3D hydrodynamical
models, \cite*{FS99} found considerable variations in the
3D-calibrated effective mixing-length parameter ($\Delta \alpha_{\rm
MLT}\sim0.10-0.15$).  When further used with the classical stellar
evolution models, these variations in the mixing-length parameter were
sufficient to produce a shift in the RGB isochrones amounting to
$\sim50$\,K or more, corresponding shift in age of $\sim30\%$ (note
that the difference between the $1$ and $15$\,Gyr isochrones in terms
of effective temperature corresponds to less than $\sim400$\,K on the
RGB/AGB -- see e.g. \cite{GBBC00}).

\begin{figure}[t]
\begin{center}
\includegraphics[width=\hsize]{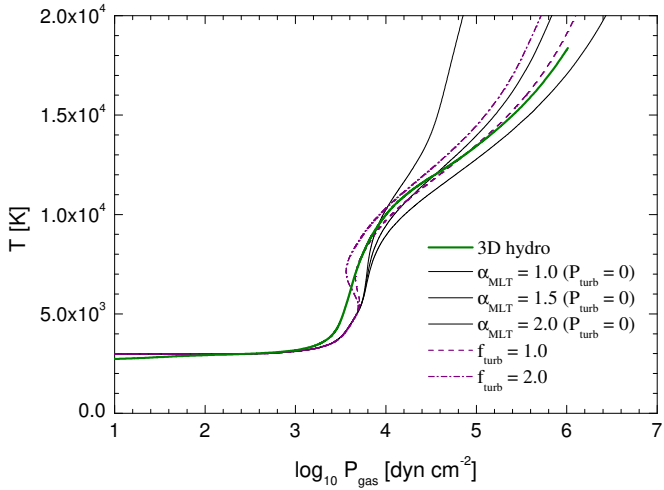}\\
\caption{Temperature stratification in the atmosphere of a red
giant star, as a function of gas pressure, $P_{\rm gas}$. The thick
solid line is a 3D hydrodynamical model (averaged on surfaces of
equal geometrical depth), thin lines are 1D plane-parallel models
calculated using different mixing-length parameter ($\alpha_{\rm
MLT}=1.0,1.5,2.0$, from left to right; turbulent pressure in all
cases equals to zero). Two 1D models with non-zero turbulent
pressure are given by dashed ($\alpha_{\rm MLT}=2.0$ and $f=1.0$)
and dashed-dotted ($\alpha_{\rm MLT}=2.0$ and $f=2.0$) lines (see
text for more details and discussion). \label{fig12}}
\end{center}
\end{figure}

Interestingly, our recent 3D hydrodynamical modelling of late-type
giant suggests that situation may be even more complicated (Ludwig et
al., in preparation). In Fig.~\ref{fig12} we show the temperature
stratification in the atmosphere of a red giant ($\Teff=3800$\,K,
$\logg=1.0$, $[\mathrm{M/H}]=0$), calculated both with conventional 1D
plane-parallel and 3D hydrodynamical model atmospheres (plotted as a
function of gas pressure, $P_{\rm gas}$). The thick solid line is
$T(P_{\rm gas})$ resulting from the 3D hydrodynamical model (averaged
on surfaces of equal geometrical depth), thin solid lines are
$T(P_{\rm gas})$ structures resulting from 1D plane-parallel models
calculated with different mixing-length parameters ($\alpha_{\rm
MLT}=1.0,1.5,2.0$, left to right). All three 1D models are calculated
assuming the turbulent pressure of $P_{\rm turb}=0$ (calculated as
$P_{\rm turb}=f \rho v^{2}$, where $\rho$ and $v$ are gas density and
convective velocity from MLT respectively, and $f$ is a dimensionless
factor, usually $f<1$). Two 1D models were calculated also with
non-zero turbulent pressure (dashed line corresponds to $\alpha_{\rm
MLT}=2.0$ and $f=1.0$ and dashed-dotted line to $\alpha_{\rm MLT}=2.0$
and $f=2.0$). This plot clearly demonstrates there may be no
reasonable choice of $\alpha_{\rm MLT}$ or $f$ available to reproduce
correctly the $T(P_{\rm gas})$ relation resulting from the 3D
hydrodynamical model within a framework of standard model atmosphere.

Clearly, attempts discussed above mark only the first steps
towards an implementation of proper treatment of convection with
an aid of 3D hydrodynamical model atmospheres. While already
interesting in itself, an extension of such a procedure to a
complete coverage of the HRD would certainly remove one of
the remaining major limitations to the predictive power of stellar
structure models.

\subsection{The realm of time-dependent phenomena}

By construction classical stellar atmospheres assume a
steady-state situation, hence cannot make predictions about
time-dependent phenomena. Fortunately, progress in observational
techniques now allows to study atmospheric processes in ever
increasing detail where time-dependence is an important aspect.
The interpretation of these observations demands for models where
time-dependence is properly accounted for.

As the first example in this category we would like to mention
dust driven winds around late-type stars, which are the result of
the interaction of stellar pulsations, dust condensation, and
radiation pressure on the formed dust grains. To obtain reliable
models it is essential to incorporate the time-dependence of
pulsations and dust formation in the models. First models able to
account for both aspects in the  time-dependent framework appeared
more then 10 years ago (\cite{FGS92}, \cite{FDH93}). The work of
\cite*{H99} marks another milestone in this ongoing development,
since here for the first time the frequency-dependence of the
radiation field was included in a dynamical wind model for an AGB
star. More recently, first attempt have been made towards a full
3D hydrodynamical modelling of an AGB star atmosphere
(\cite{FH03}).

\begin{figure}[t]
\begin{center}
\includegraphics[width=0.49\hsize]{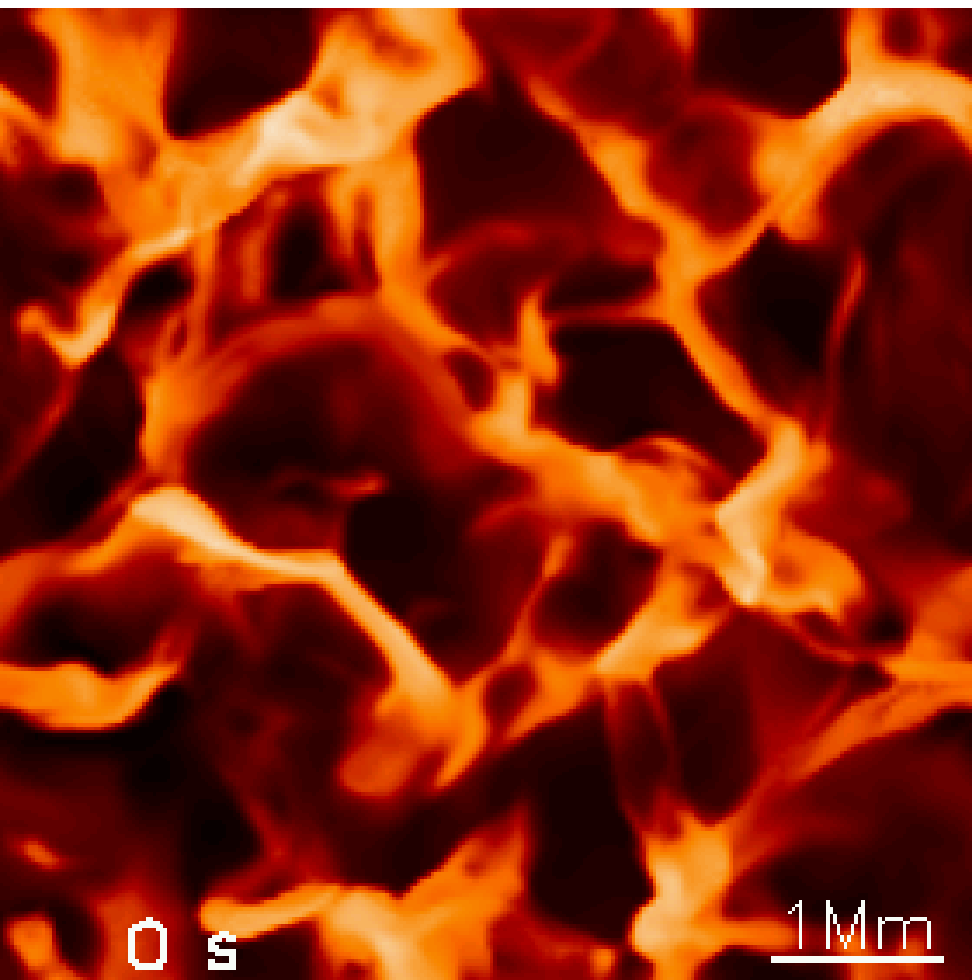}
\includegraphics[width=0.49\hsize]{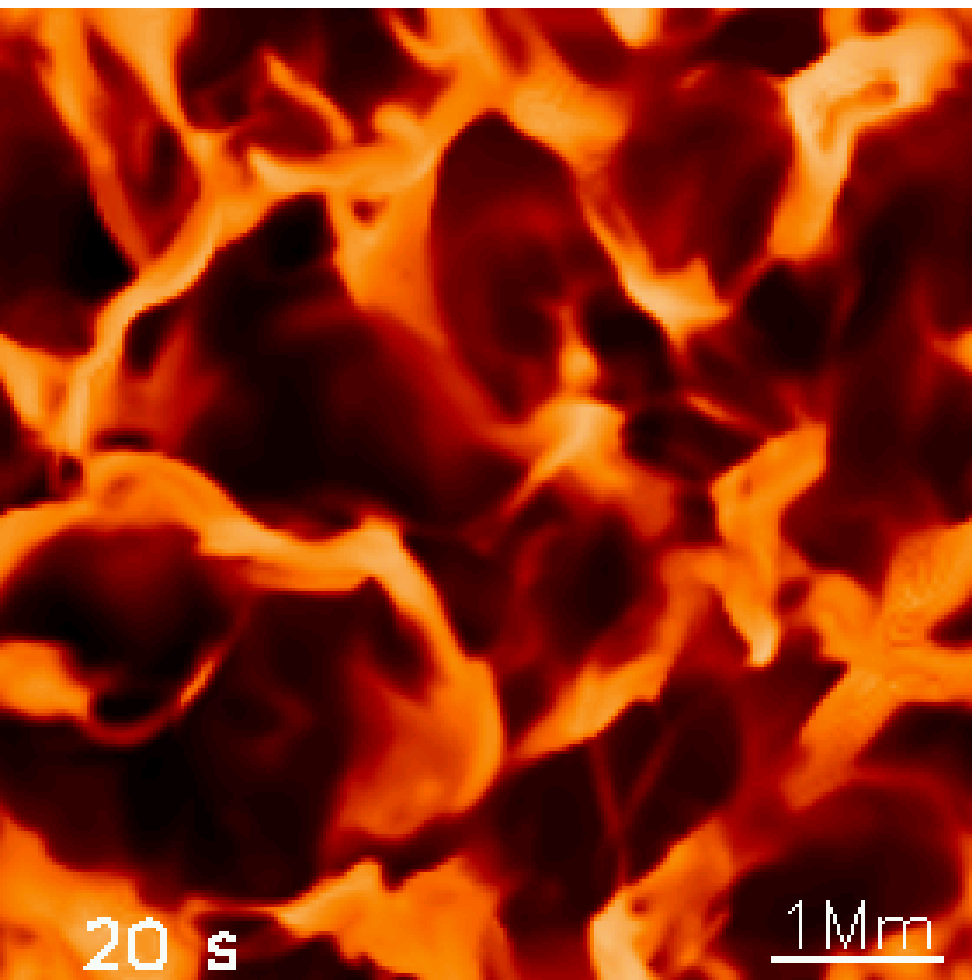}\\[\baselineskip]
\caption{Small scale chromospheric (shock) structures as (potentially)
  seen in the 1\pun{mm} radio continuum (cf. Wedemeyer-B\"ohm et al.,
  this volume)\label{fig2}}
\end{center}
\end{figure}

Another area where the inclusion of time-dependent processes is
essential is the structure of stellar chromospheres and coronae.
Their heating is provided by the dissipation of acoustic or
magneto-hydrodynamic waves, as well as magnetic reconnection
driven by forces on the magnetic field lines exerted by fluid
motions. Historically, so called semi-empirical atmosphere models
have been constructed to reproduce observed spectral features.
While by construction being able to reproduce the particular
observations, they lack the ability to rigorously test our
physical understanding, and also lack the power to predict
atmospheric properties in a regime of stellar parameter distinct
from the very ones they have been adjusted to.  A prominent
example of the impact of a hydrodynamical model atmosphere in this
are is the work of \cite*{CS95} who constructed a one-dimensional
time-dependent model of of the solar chromosphere. As a
consequence of their modelling efforts they proposed a shift of
the paradigm how to describe stellar chromospheres in general.
Recently, \cite*{WFSLH04} addressed the same problem with a
three-di\-men\-sio\-nal hydrodynamical model. The detailed spatial
information the model provides can be exploited for the prediction
of otherwise unavailable properties. Figure~\ref{fig2} shows a
predicted image of the structure of the solar chromosphere as seen
in the 1\pun{mm} radio continuum as potentially observable by the
Atacama Large Millimeter Array (ALMA).

Perhaps now stretching a bit the notion of what a model atmosphere
should actually model we would also like to mention an application of
hydrodynamical model atmospheres which at present gets increasingly
interesting due to dedicated satellite missions: the excitation of
solar-like oscillations. It is believed that the solar 5\pun{min}
oscillations are excited by the ``noise'' generated by the stochastic
granular gas motions. \cite*{NS01} and \cite*{SN01} have demonstrated
for the case of the Sun that hydrodynamical model atmospheres can
provide an estimate of the power which is injected into the
oscillations by the granular flow field. The advantage of the method
is that the description of the stochastic flow in is quite realistic
while in analytical approaches one has to rely on approximations of
the flow properties whose validity are difficult to judge. Recently,
\cite{SGTLN04} have applied the method to a broader range of stars.

\begin{figure}[t]
\begin{center}
\includegraphics[width=0.8\hsize]{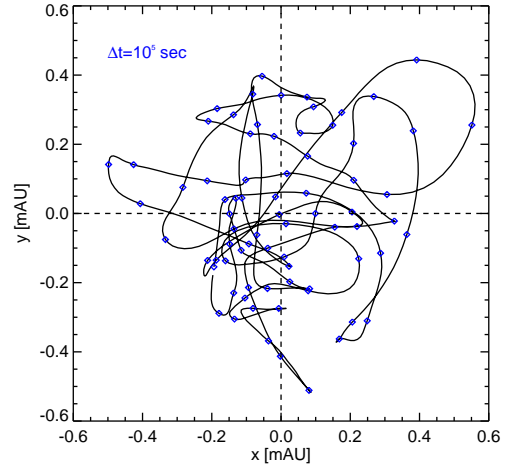}\\
\caption{Path of the photometric centroid of a red giant star (cf.
 Svensson \&\ Ludwig, this volume).\label{fig4}}
\end{center}
\end{figure}

Related to the above issue is the question of the ultimate photometric
stability, the stability of the photo-centric position, and the
stability of the spectroscopically derived radial velocities of stars
with a time-variable granular surface pattern. Hydrodynamical model
atmospheres can predict the convectively induced stellar
variability. Fig.~\ref{fig4} shows an example of the predicted motion
of the apparent position (the photometric centroid) of a red giant in
the plane of the sky.

\section{Future developments: theory \&\ observations}

In this section we want to give some projections of the future in
terms of model developments, model results, and observations which aid
model developments.
We also want to explain why extended grids of
hydrodynamical including spectral diagnostics have not been available
so far.

\subsection{Computational challenges}

Hydrodynamical model atmospheres are often perceived as
computationally ``heavy weight'' --- as a supercomputing
application. Today however, such a notion is not fully justified
as long as we are considering ``typical'' hydrodynamical models,
i.e. models of modest spatial (about 100 grid points per
dimension, meaning in total $10^6$ points) and wavelength
resolution (4--7 equivalent wavelength points). Despite the
obvious limitations in spatial and wavelength resolution such
models describe the atmospheric dynamics with a high degree of
fidelity (see e.g. \cite{ALNS00}). Of course, hydrodynamical
models are being improved in terms of spatial and wavelength
resolution. But progress in this respect is rather slow and mostly
parallels the increase of computer performance since the
computational costs scale as the forth power of the spatial
resolution per dimension (linearly with the total number of grid
points, and one power due to a related linear decrease of the time
step), and linearly with the number of employed wavelength points.
High resolution models in fact still demand for supercomputing
resources.

However, being content to reproduce spectral diagnostics with typical
hydrodynamical models, such a model takes $\approx 1$ month on PC-type
machine to compute in the case of the Sun. Due to tighter constraints
on the time stepping related to the radiative transfer red giants and
A-type stars need $\approx 10\ldots 100$ times longer computing times.
All in all this is not so bad since low-cost PC-clusters or fast
parallel machines are now commonly available. So, what keeps
modelists from providing grids of hydrodynamical model together with
spectral diagnostics like flux distributions and colours? Two stumbling
blocks have prevented this.

Stumbling block 1: the calculation of the spectral diagnostics is
computationally very demanding. The spectrum synthesis calculation
of a spectrum with about $10^5$ wavelength points for a 1D
atmosphere takes $\approx 10\pun{PCmin}$. The full
spatial-temporal information of a typical hydrodynamical model
atmosphere corresponds to $\approx 10^6$ equivalent 1D
atmospheres. The given computing time for a synthetic spectrum
refers to calculations under the assumption of LTE. Non-LTE aspects
might increase this number substantially. Obviously, at the moment
tackling the spectral synthesis problem head on does not work.
However, this is perhaps not necessary since the full
spatial-temporal information is not needed, but mostly time and
spatial (i.e. stellar disk integrated) averages. It is likely that
one can come up with certain ``short-cuts'' which allow a more
economic evaluation of the overall spectral characteristics. The
development of such procedures is pending.

Stumbling block 2: there is an increased demand on the computing
logistics. The operation of hydrodynamical model atmosphere codes
is not fully automated. Manual intervention starts with the
construction of reasonable initial models. Even hydrodynamics
codes which are considered as robust demand for initial conditions
which are physically not too far apart from actual conditions.
Hence, a certain degree of (human) creativity is needed for
constructing initial models for atmospheric parameters not studied
before.  Moreover, large amounts of numerical data are generated
during a hydrodynamical model run. While storage of data of a
single run does not pose a problem, the data of a larger number of
runs usually demand for disk-array-capacity of storage. The output
undergoes analysis steps producing spectral information and
further auxiliary diagnostics. A fully automatic end-to-end
organisation of the process has not been achieved yet, though it
will be indeed important when simultaneously handling of the order
of hundred or more models.

The stumbling blocks described before are not of principal nature, and
will be overcome. The steady increase of computing power and size of
available storage space provide almost automatically alleviates the
situation. Perhaps the present bottleneck for the speed of development
is the available woman- or man-power scattered out over the few
involved working groups. Nevertheless, we believe that in the
foreseeable future grids of hydrodynamical models including some
observational diagnostics will become available.

\subsection{Observations aiding model developments}

What kind of observations are the modelist waiting for which would
aid model development and validation? Similar to classical model
atmospheres ``1D like'' observables can be used the check the
validity of hydrodynamical models as well. Precise measurements of
the center-to-limb variation for different kinds of stars can be
used to test the thermal structure of a model, i.e. its average
thermal profile and thermal fluctuations. Spatially unresolved,
high resolution spectra allowing to measure spectral line shifts
and bisectors provide statistical constraints on the velocity
field combined with the temperature fluctuations. The measurement
of stellar radii puts constraints on the temperature structure of
the convective envelope, allowing to test model predictions of the
mixing-length parameter. The observations listed above are partly
already available, and have been used in the validation of
hydrodynamical models. They are particularly useful if combined
with independent supplementary information concerning the global
stellar structure, like the stellar mass, luminosity, and chemical
composition.

Beyond the classical observables, ``multi-D'' observables (i.e.
those specifically related to spatial or temporal inhomogeneity)
can provide further useful additional constraints. They are now in
reach or going to get into reach due to refined observing
techniques, or newly established techniques like optical
interferometry. The direct measurement of convective scales and
their brightness contrast would provide a direct test of
hydrodynamical model atmospheres. This could be combined with the
time-wise statistics of the granulation pattern. The time domain
could also be more indirectly addressed by observing the result of
macroscopic transport or mixing processes competing with other
processes. An example would be deviations from equilibrium
chemistry due to the competition between transport and formation
of a particular chemical constituent.  A specific case is the
distribution of condensates (clouds) in the atmospheres of very
cool stars, brown dwarfs, and planets.

\section{Conclusions \&\ final remarks}

We hope that we convinced the reader that most convective stellar
atmospheres are now accessible to 3D hydrodynamical modelling. 3D
hydrodynamical model atmospheres have been already been proven
useful in the development of our theoretical understanding of cool
star stellar atmospheres in general, and the interpretation of
observations. Their ``esoteric'' character is diminishing, as
methodological developments, as well as the steady increase of
computing power, have been and are transforming them more and more
into a standard tool for stellar physics. While not quite there
yet, we believe that grids of hydrodynamical model atmospheres
including some spectral diagnostics will become available in the
foreseeable future. Being aware that prediction are hard to make
--- especially when it comes to the future of such rapidly evolving
field of modern astrophysics --- we think that this is going to
happen within five years from now.  Observers who are not that
patient and think that their work could benefit from input from
predictions from hydrodynamical models are encouraged to seek
collaborations with model developers. As welcomed effect of such
collaborations further model developments could be spawned since
models would have to be adapted to the particular observational
needs.

\end{document}